\documentclass{emulateapj_mo}
\slugcomment{} 
\shorttitle{SDSS Quasar Lens Search. I.}
\shortauthors{Oguri et al.}
\begin{document}
\title{The Sloan Digital Sky Survey Quasar Lens Search. I. \\
Candidate Selection Algorithm}  
%
\author{
Masamune Oguri,\altaffilmark{1}
Naohisa Inada,\altaffilmark{2}
Bartosz Pindor,\altaffilmark{3}
Michael A. Strauss,\altaffilmark{1}
Gordon T. Richards,\altaffilmark{4}\\
Joseph F. Hennawi,\altaffilmark{5}
Edwin L. Turner,\altaffilmark{1}
Robert H. Lupton,\altaffilmark{1}
Donald P. Schneider,\altaffilmark{6}\\
Masataka Fukugita,\altaffilmark{7,8}
and
Jon Brinkmann\altaffilmark{9}
}
\altaffiltext{1}{Princeton University Observatory, Peyton Hall,
Princeton, NJ 08544.}
\altaffiltext{2}{Institute of Astronomy, Faculty of Science, 
University of Tokyo, 2-21-1 Osawa, Mitaka, Tokyo 181-0015, Japan.}
\altaffiltext{3}{Department of Astronomy, University of Toronto, 60
St. George Street, Toronto, Ontario M5S 3H8, Canada.}  
\altaffiltext{4}{Department of Physics and Astronomy, Johns Hopkins
University, 3701, San Martin Drive, Baltimore, MD 21218.}
\altaffiltext{5}{Department of Astronomy, University of California at
Berkeley, 601 Campbell Hall, Berkeley, CA 94720-3411.} 
\altaffiltext{6}{Department of Astronomy and Astrophysics, The
  Pennsylvania State University, 525 Davey Laboratory, University
  Park, PA 16802.}  
\altaffiltext{7}{Institute for Advanced Study, Einstein Drive,
  Princeton, NJ 08540.} 
\altaffiltext{8}{Institute for Cosmic Ray Research, University of
  Tokyo, 5-1-5 Kashiwa, Kashiwa City, Chiba 277-8582, Japan.} 
\altaffiltext{9}{Apache Point Observatory, P.O. Box 59, Sunspot, NM88349.}
%
%
\begin{abstract}
We present an algorithm for selecting an uniform sample of
gravitationally lensed quasar candidates from low-redshift
($0.6<z<2.2$) quasars brighter than $i=19.1$ that have been
spectroscopically identified in the Sloan Digital Sky Survey
(SDSS). Our algorithm uses morphological and color selections
that are intended to identify small- and large-separation lenses,
respectively. Our selection algorithm only relies on parameters that
the SDSS standard image processing pipeline generates, allowing easy
and fast selection of lens candidates. The algorithm has been tested
against simulated SDSS images, which adopt distributions of field and
quasar parameters taken from the real SDSS data as input. Furthermore,
we take differential reddening into account. We find that our selection
algorithm is almost complete down to separations of $1''$ and flux
ratios of $10^{-0.5}$. The algorithm selects both double and quadruple
lenses. At a separation of $2''$, doubles and quads are selected with
similar completeness, and above (below) $2''$ the selection of quads
is better (worse) than for doubles. Our morphological selection
identifies a non-negligible fraction of single quasars: To remove
these we fit images of candidates with a model of two point sources
and reject those with unusually small image separations and/or large
magnitude differences between the two point sources. We estimate the
efficiency of our selection algorithm to be at least 8\% at image
separations smaller than $2''$, comparable to that of radio
surveys. The efficiency declines as the image separation increases,
because of larger contamination from stars. We also present the
magnification factor of lensed images as a function of the image
separation, which is needed for accurate computation of magnification
bias. 
\end{abstract}
 
 
\section{Introduction}

Gravitational lensing is a unique tool for exploring cosmology and the 
structures of astronomical objects. The well-understood physics of
lensing makes it straightforward to use as a cosmological
probe. Moreover, it is the only method that probes distributions of dark
matter directly, since gravitational lensing is a purely gravitational
phenomenon. 

While detailed investigations of single intriguing lens systems are
useful for cosmological and astrophysical applications, for some
studies it is essential to do statistical analyses of a complete
sample of lenses that are selected from a well-understood source
population. For instance, strong lens probabilities are sensitive to
the volume element of the universe, and thus the amplitude of the
cosmological constant, thus we can constrain cosmological models by
comparing the observed lensed fraction in a catalog of distant sources
to theoretical expectations
\citep{turner90,fukugita90,kochanek96,chiba99}. Strong lensing
probabilities at the cluster mass scale probe the abundance and mass
distribution of clusters \citep{narayan88,maoz97}. The lens image
separation distribution from galaxy to cluster mass scales can be used  
to study baryon cooling in dark halos \citep{kochanek01} and the
connection between galaxy luminosities and masses of host halos
\citep{oguri06}. 

A statistical sample is important not only for studies of lensing
rates, but also for interpretations of the results of individual lens
modeling. It has been argued that lensing objects may be more or less
atypical in  their shapes, orientations \citep{hennawi06b}, and
environments \citep{oguri05b} because of strong dependences of lensing
probabilities on these quantities. Therefore we  must take proper
account of lensing biases to make a fair comparison between theory
and observed results, which can only be accomplished for a well-defined
statistical lens sample. A similar argument holds for constraints
from stacked samples of strong lenses, including the structure and
evolution of early-type galaxies \citep{treu04,rusin05} and the
fraction of substructures in lens galaxies \citep{dalal02}. It is
quite hard to estimate the degree of the lensing biases if one adopts
a heterogeneous sample of lenses that have been discovered in a
variety of ways. 

The current largest complete sample of strong lenses was selected 
from a radio survey: The Cosmic-Lens All Sky Survey
\citep[CLASS;][]{myers03,browne03} has discovered 22 gravitational
lenses from $\sim$16,000 radio sources. 13 of these lenses out of 
$\sim$9,000 flat-spectrum radio sources constitute a statistically
well-defined sample. Recent progress on large-scale optical
surveys has made it possible to construct comparable or even larger lens
samples in the optical band. In particular, the Sloan Digital Sky Survey
\citep[SDSS;][]{york00} is expected to identify $\sim$100,000
confirmed quasars \citep{schneider05} and $\sim$500,000 photometrically
selected quasars \citep{richards04} when completed. In addition to the
large number of sources, the optical lens survey has the advantage of a
source population whose redshift and magnitude distributions are well
understood, which is important when doing statistical analyses. On the
other hand, optical lensed quasars are sometimes significantly
affected by the light from foreground lens galaxies and dust
reddening. Therefore, optical and radio lens surveys contain different
systematics, and it is thus of great importance to compare the results
from these two complementary lens samples.  

We are searching for optical gravitationally lensed quasars from the
imaging and spectroscopic data of the SDSS to construct a statistical
sample of lensed quasars.  Indeed, this project, the SDSS Quasar Lens
Search (SQLS), has already discovered 14 new lensed quasars, and has
recovered several ($\sim 7$) previously known lensed quasars
\citep*{inada03a,inada03b,inada03c,inada05,inada06,johnston03,morgan03,
pindor04,pindor06,oguri04b,oguri05a,burles06}. Although these
lenses do not constitute a statistically well-defined sample, these
discoveries give promise that we can create a large complete sample of
lensed quasars. In order to construct a statistical sample, we need
(i) to select lens candidates by a homogeneous method, and (ii) to
understand the completeness of the method. In reality, we must also
keep the efficiency reasonably high to assure the feasibility of 
follow-up observations. We can take advantage of the deep SDSS imaging
data in five broad bands for the efficient identification of lensed
quasar candidates. \citet[][hereafter P03]{pindor03} presented a
selection method for lensed quasars from the SDSS, and quantified its
selection function. Their selection algorithm compares the chi-square
value of a two-component point-spread function (PSF) model of
spectroscopically confirmed SDSS quasars with that of a one-component
PSF model. From simulated SDSS images, they found that the algorithm
recovers quasar pairs of similar fluxes down to separations of $\sim
0\farcs7$. 

In this paper, we present another efficient algorithm to identify
low-redshift ($z\lesssim 2.2$) lensed quasar candidates from the
spectroscopic sample of SDSS quasars. Our selection algorithm uses
both morphological and color selection. The former aims to select
quasars that are extended and are not well fitted by a PSF; it is 
designed to identify small-separation ($\sim 1''$) lens candidates.
The latter algorithm simply compares colors of close pairs of objects
with those of source quasars, and is focused on lenses whose image
separations are large enough to be deblended in the SDSS imaging data
reduction pipeline. Our selection algorithm relies only on the
outputs of the SDSS imaging pipeline and therefore does not  require
independent analysis of the imaging data themselves. This allows easy
and fast identification of lensed quasar candidates.  We apply our
algorithm to realistically simulated SDSS images to quantify the
completeness in detail. For the simulation of lenses, we follow
the methodology developed by P03. We adopt observed distributions of
sky levels and seeing sizes, as well as the observed SDSS source
quasar population in a real SDSS quasar catalog, to estimate the
completeness of our algorithm in a realistic manner. The algorithm
will be applied to the SDSS data in Paper II (N. Inada et al., in
preparation).

This paper is organized as follows. In \S \ref{sec:source}, we discuss
the source quasar sample suitable for lens statistics. Our lens
selection algorithms are presented in \S \ref{sec:sel}, and are tested
against simulations in \S \ref{sec:sim}. Section \ref{sec:second} is
devoted to study an additional selection that is intended to remove any
point-source like candidates. The efficiency of our algorithms is
briefly discussed in \S \ref{sec:eff}, and we summarize our results
in \S \ref{sec:con}.

\section{Source Quasar Sample}
\label{sec:source}

The SDSS is a survey to image $10^4$ ${\rm deg^2}$ of the sky, and
also to conduct spectroscopy of galaxies, quasars, and stars that are
selected from the imaging data
\citep{eisenstein01,strauss02,richards02}. The tiling algorithm
\citep{blanton03} assigns targets to fibers.\footnote{Because of the
physical size of fibers, spectra of no two targets that are closer
than $55''$ are taken on a single plate. This ``fiber collision''
limitation makes it unlikely to obtain spectra of more than one
component in a gravitational lens.} A dedicated 2.5-meter telescope
\citep{gunn06} at Apache Point Observatory (APO) is equipped with a
multi-CCD camera \citep{gunn98} with five optical broad bands centered
at $3551$, $4686$, $6166$, $7480$, and $8932${\,\AA}
\citep{fukugita96,stoughton02}.  The imaging data are automatically
reduced by the photometric pipeline PHOTO \citep{lupton01}. The
astrometric positions are accurate to about $0\farcs1$ for sources
brighter than $r=20.5$ \citep{pier03}, and the photometric calibration
errors are typically less than 0.03 magnitude
\citep{hogg01,smith02,ivezic04}. All data are being released on a
regular schedule
\citep{stoughton02,abazajian03,abazajian04,abazajian05,adelman06}.  

We construct a statistical sample of lensed quasars from the spectroscopic
SDSS quasar catalog \citep{schneider02,schneider03,schneider05}. The
quasar sample is quite heterogeneous in the sense that the quasars are
selected with several different methods and the completeness is
strongly dependent on the properties of quasars \citep[e.g., magnitudes, 
redshifts, morphologies; see][]{richards06b}. We make a subsample of
source quasars, which are suitable for lens statistics, by posing the
following restrictions.

\begin{figure}
\epsscale{0.98}
\plotone{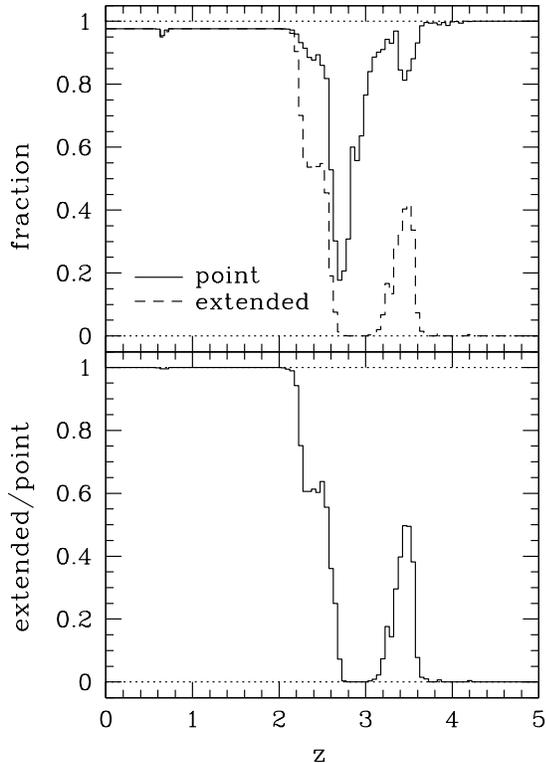}
\caption{The completeness of the quasar spectroscopic target selection for
  point and extended quasars, computed from the simulation done by
  \citet{richards06b}. We show completeness averaged over $15<i<19.1$. 
  The ratio of the two completenesses is shown in the lower panel. 
\label{fig:compl}}
\end{figure}

\begin{enumerate}
 \item We consider only quasars in the redshift range $0.6\leq 
 z\leq 2.2$. Lower-redshift quasars are often extended in the SDSS
 images, which makes the morphological selection (\S
 \ref{sec:sel_mor}) quite difficult; in any case, these quasars have
 low lensing probabilities. At low redshift most quasars are targeted
 by UV excess which can select both point and extended quasars
 efficiently, while at higher redshifts outliers from the $griz$
 stellar locus are targeted as quasar candidates, for which we had to
 restrict our attention to point sources because of large
 contamination from galaxies. This is clearly shown in Figure
 \ref{fig:compl}, which plots the completeness obtained from the
 simulation by \citet{richards06b}. The completenesses are different
 at $z\geq 2.2$ between point and extended quasars, which introduces a
 non-negligible anti-lens bias. In particular, extended quasars at
 $z\gtrsim 3$ are not selected by the quasar target selection
 algorithm, which significantly reduces the number of $\sim 1''$
 separation high-redshift lenses in the SDSS spectroscopic quasar
 catalog (see also Appendix \ref{sec:append}). 
 In addition, at $z>2.2$ it becomes more difficult to use colors to
 distinguish quasars from stars, thus the efficiency of lens candidate
 selections is significantly degraded, particularly at the
 large-separation region  where contamination from stars are quite large. 
\item We adopt quasars with Galactic extinction corrected
  \citep{schlegel98} $i$-band magnitudes $15.0\leq i_{\rm cor}\leq
  19.1$. This is because at $i_{\rm cor}\leq 19.1$ the low-redshift
  ($z\lesssim 2.2$) quasar target selection is almost complete
  \citep{richards02,richards06b}, while fainter quasars are selected
  mainly with various ``serendipity'' algorithms \citep{stoughton02}
  which are very incomplete and may introduce biases in estimating
  lensing rates. Moreover, the faint quasars have several disadvantages,
  including large photometric errors that reduce the efficiency
  of lens selection algorithms, and contamination by lens galaxies
  that become more important for less luminous quasars
 \citep[e.g.,][]{kochanek96}. Fainter lenses are identified as lens
  candidates with lower efficiency because of their lower
  signal-to-noise ratios (see \S \ref{sec:compl}).
 \item Close lens pairs are difficult to distinguish from single PSFs
 if the seeing is poor. Therefore we reject quasars in fields if the
 $i$-band image has the SDSS imaging parameter {\tt PSF\_WIDTH}
 (effective PSF width based on 2-Gaussian fit) $>1\farcs8$. This
 criterion rejects roughly $2\%$ of quasars.   
\end{enumerate}
  
As a specific example, we adopt the 46,420 quasars in the DR3 quasar
sample \citep{schneider05} as our parent sample. By restricting the
range of $i$-band magnitudes and redshifts, the number of quasars
decreases to 23,316 ($\sim 50\%$). The final number of quasars, after
rejecting poor seeing fields, is 22,682 ($\sim 49\%$).

\section{Selection Algorithm}
\label{sec:sel}

In this section, we present our two selection algorithms, using
morphological and color selection criteria, respectively. The
algorithms are intended to identify low-redshift ($z\lesssim 2.2$)
lensed quasars, though their extension to higher-redshift lenses is 
rather straightforward (besides the inefficiency and imcompleteness of
high-redshift lenses). Lens candidates are identified using imaging
data; we do not use any spectroscopic information, although in some
cases spectra offer a powerful way to locate gravitational lensing 
\citep[e.g.,][]{johnston03,bolton05,bolton06}.
We determine the selection criteria described below mostly in an
empirical manner; we choose criteria to keep the completeness high,
for both known lenses and simulations done in \S \ref{sec:sim}, and at
the same time have reasonable efficiency when applied to the SDSS
data.   

\subsection{Morphological Selection}
\label{sec:sel_mor}

When the image separation between multiple images is small
($\theta\lesssim 2\farcs5$), PHOTO cannot deblend the system into two 
components, and the objects are classified as single extended objects.
Thus small-separation lenses can be identified as lens candidates by
searching for objects that are poorly fitted with the local PSF. 
To do so, we use the following two SDSS imaging parameters.  One is
{\tt objc\_type} which describes the classification of the object as
stars or galaxies, determined from the differences of magnitudes
obtained by fitting PSF and galaxy profiles to the images in each band
\citep[e.g.,][]{stoughton02};  {\tt objc\_type} $=6$ indicates the
object is a point source, while {\tt objc\_type} $=3$ means the object is
extended. The other parameter, {\tt star\_L} (available in each band),
is the probability that an object would have at least the measured
value of $\chi^2$ of a fit of the image to the PSF, if it really is
well represented by a PSF. Put another way, the higher the value of
{\tt star\_L}, the more likely it is fitted by a PSF.  
   
The specific selection criteria of this morphology selection are as
follows. First, even if a quasar is classified as a point source, we
select it as a lens candidate if the PSF likelihood is
small. Specifically we select objects that satisfy all of the
following four conditions:
\begin{eqnarray}
\mbox{(M1):} && {\tt objc\_type}=6, \nonumber\\
&& {\tt star\_L}(u)\leq 0.03, \nonumber\\
&& {\tt star\_L}(g)\leq 0.04, \nonumber\\
&& {\tt star\_L}(r)\leq 0.07\;\;{\rm OR}\;\;{\tt star\_L}(i)\leq 0.07.
\end{eqnarray}
We do not use the $z$-band {\tt star\_L} parameter because of the
relatively low signal-to-noise ratio (S/N) of $z$-band
images. Although the S/N of $u$-band images are also low, we use the
$u$-band parameter because quasars are UV-excess sources and thus the
$u$-band is quite effective in discriminating quasar-quasar pairs from
quasar-star pairs. However, we find that the condition on the $u$-band 
parameter is sometimes too strong, resulting in missing some lens
candidates. As a separate cut, we adopt a relaxed range of $u$-band
star likelihood, but slightly tighter conditions on other
parameters:  
\begin{eqnarray}
\mbox{(M2):} && {\tt objc\_type}=6, \nonumber\\
&& {\tt star\_L}(u)\leq 0.06, \nonumber\\
&& {\tt star\_L}(g)\leq 0.04, \nonumber\\
&& {\tt star\_L}(r)\leq 0.04\;\;{\rm OR}\;\;{\tt star\_L}(i)\leq 0.04.
\end{eqnarray}
When a quasar image is classified as extended, we use a similar but
relaxed, criterion to identify a lens candidate; an object is selected
when all the following three conditions are met:
\begin{eqnarray}
\mbox{(M3):} && {\tt objc\_type}=3, \nonumber\\
&& {\tt star\_L}(u)\leq 0.45\;\;{\rm OR}\;\;{\tt star\_L}(g)\leq 0.35,
 \nonumber\\ 
&& {\tt star\_L}(r)\leq 0.60\;\;{\rm OR}\;\;{\tt star\_L}(i)\leq 0.60.
\end{eqnarray}
We select quasars that satisfy one or more of the criteria (M1)-(M3) as
lens candidates. In practice, approximately half of small-separation
gravitational lenses are classified as extended objects and selected 
by the condition (M3). Lenses that are classified as point-sources  
tend to satisfy both conditions (M1) and (M2); the difference
between (M1) and (M2) becomes important only for limiting cases in which
lenses are close to the magnitude limit ($i\sim19.1$) and/or have very
small image separations ($\theta\lesssim1''$). These conditions are
designed using both real SDSS data and simulations presented in \S
\ref{sec:sim}. 

\subsection{Color Selection}
\label{sec:sel_col}

When the image separation is large enough for PHOTO to deblend two
components, we can select lens candidates by comparing the colors between
spectroscopically confirmed quasars and candidate companions. At
$\theta>7''$ the candidates' companions must be point sources, but for
smaller separations we also allow extended sources to be candidate
companions, because for such small separations lensing galaxies may be
superposed on the fainter components (e.g., Q 0957+561).  The selection
method presented here is somewhat similar to those given by
\citet{oguri04a} and \citet{hennawi06a}, but we follow P03 to extend the
methods to select differentially reddened lenses. 

\begin{figure*}
\epsscale{0.65}
\plotone{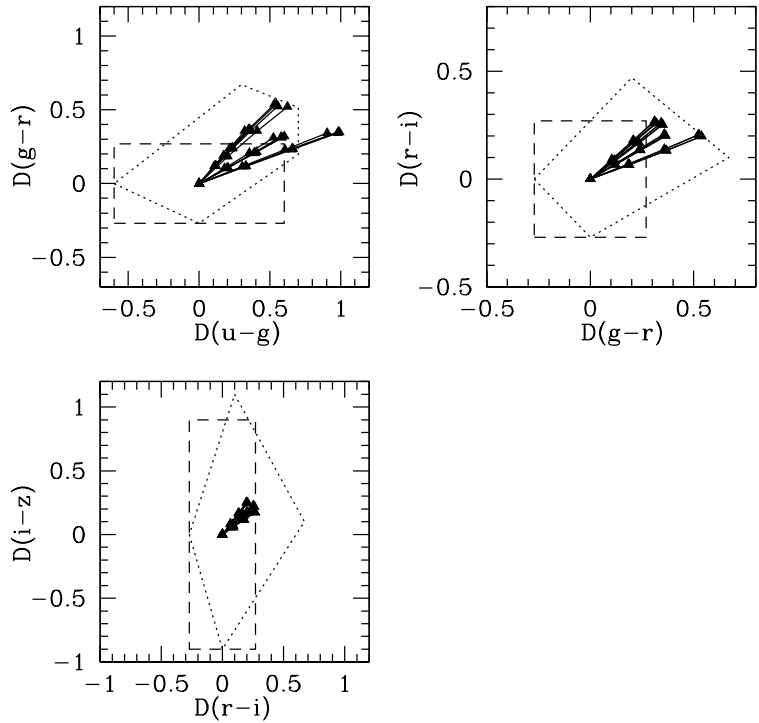}
\caption{Reddening vectors in color difference-color difference space
 are shown by solid lines plus filled triangles. We compute the vectors
 for the quasar redshift $z_s=0.7$, $1.2$, $1.7$, $2.2$, and the lens
 redshift $z_s=0.2$, $0.4$, $0.6$, $0.8$. The differential extinction
 $\Delta E(B-V)$ is changed from 0 to 0.3 (triangles are plotted every
 0.1 increment). The color selections (C1) and (C2) are shown by
 dashed and dotted boxes, respectively.
\label{fig:red}}
\end{figure*}

We conduct the color selection on the basis of the color difference:
\begin{eqnarray}
D(i-j)&=&(i-j)_{\rm quasar}-(i-j)_{\rm companion} \nonumber\\
&& \;\;{\rm
  OR}\;\; (i-j)_{\rm companion}-(i-j)_{\rm quasar},
\end{eqnarray}
where ($i$, $j$)$=$($u$, $g$), ($g$, $r$), ($r$, $i$), ($i$, $z$), and
quasar and companion indicate the SDSS spectroscopic quasar in the source
quasar sample and candidate companion, respectively. We use PSF
magnitudes and their errors throughout the paper. 
Although the definition of $D(i-j)$ contains a sign ambiguity,
we regard candidate pairs as lens candidates if they pass
the criteria below using either one of the above definitions: This becomes
important when we consider differential reddening, because at
first sight we do not know which object is reddened more. When the color
difference comes from reddening, $D(i-j)$ for all set of ($i$, $j$)
must have the same sign. On the other hand, the color difference of
two unrelated objects could have both positive and negative
signs for different ($i$, $j$). This is the reason we do not define
$D(i-j)$ as the absolute value of the color difference. 

Since gravitational lensing does not change the color of objects, the
color difference should be small if two components are lensed
images of a single quasar. This criterion can be written as follows:
\begin{eqnarray}
\mbox{(C1):} &&|D(i-j)|<\Delta D(i-j),
\end{eqnarray}
which must be met for all four colors. The limit $\Delta D(i-j)$
describes an acceptable error of the color difference, which we adopt
the following fixed values 
\begin{eqnarray}
\Delta D(u-g)&=&3 \times 0.20,\nonumber\\
\Delta D(g-r)&=&3 \times 0.09,\nonumber\\
\Delta D(r-i)&=&3 \times 0.09,\nonumber\\
\Delta D(i-z)&=&3 \times 0.30.
\label{eq:colordif_er}
\end{eqnarray}
These values correspond to typical 3-$\sigma$ errors of the color
differences of faint ($\sim19-20$ mag) objects in the SDSS images. We
find that color differences contain very large errors when 
image separations are small because of the SDSS deblending
algorithm. To account for this, we replace $\Delta D(i-j)$ with
$3\times \sigma_{D(i-j)}$ for $2''\leq\theta<3\farcs5$, and with
$6\times \sigma_{D(i-j)}$ for $\theta\leq 2''$, where
$\sigma_{D(i-j)}$ is defined by 
\begin{equation}
\sigma_{D(i-j)}=\sqrt{(\sigma_i^2+\sigma_j^2)_{\rm quasar}+
(\sigma_i^2+\sigma_j^2)_{\rm companion}},
\end{equation}
where $\sigma_i$ are PSF magnitude errors estimated by PHOTO. 
Since it is quite rare that objects lie at so close
to the quasar, the large errors have little effect on the
efficiency of our lens selection algorithm. 

While gravitational lensing is independent of wavelength, significant
color differences between lensed components may be caused by
reddening. For instance, \citet{falco99} measured the distribution of 
differential reddening $\Delta E(B-V)$ from a number of
gravitationally lensed quasar systems and found that is well described
by a Gaussian with a zero mean and standard deviation of $0.1$ mag.  We
estimate how the differential reddening affects the color difference
$D(i-j)$ as follows (see also P03). First, we use the composite
quasar spectrum of \citet{vandenberk01} as an input quasar spectral 
energy distribution. Next we adopt the Galactic extinction 
law\footnote{We note that our result is not strongly dependent on the
  assumed reddening law. For instance, we have repeated the analyses
  adopting SMC-like reddening, which \citet{hopkins04} claims fits the
  observed reddening of quasars, and confirmed that the
  difference of completeness (\S \ref{sec:compl}) is indeed small.} of
\citet{cardelli89} and compute the color difference as a function of
quasar and lens redshifts and $\Delta E(B-V)$, assuming $R_V=3.1$. We
show estimated reddening vectors for several sets of quasar and lens
redshifts in Figure \ref{fig:red}. As expected, the effect of the
differential reddening is more significant for bluer bands. From this
estimation, we prepare another color selection algorithm in
color-color space that is designed to select lens systems with large
differential reddening:  
\begin{eqnarray}
\mbox{(C2):}&& y>-\frac{\Delta y}{\Delta x}x-\Delta y, \nonumber\\
&& y<\frac{y_2-y_1-\Delta y}{x_2-x_1+\Delta x}(x-x_1)+y_1+\Delta y,
 \nonumber\\ 
&& y>\frac{y_2+\Delta y}{x_2+\Delta x}x-\Delta y, \nonumber\\
&& y<\frac{y_1+\Delta y}{x_1+\Delta x}(x+\Delta x),\nonumber\\
&& x<0.2+\Delta x\;\;\;\mbox{IF}\;\; x=D(u-g),
\end{eqnarray}
where the condition must be satisfied for all set of ($x$,
$y$)$=$($D(u-g)$, $D(g-r)$), ($D(g-r)$, $D(r-i)$), ($D(r-i)$, $D(i-z)$),
and  
\begin{eqnarray}
  (x_1,y_1,x_2,y_2) &&\nonumber\\
&&\hspace*{-24mm}=\left\{
      \begin{array}{ll}
        (0.3,0.4,0.4,0.4) & (x,y)=(D(u-g),D(g-r)), \\ 
	(0.2,0.2,0.4,0.1) & (x,y)=(D(g-r),D(r-i)), \\ 
	(0.1,0.2,0.4,0.1) & (x,y)=(D(r-i),D(i-z)).  
      \end{array}
   \right. 
\end{eqnarray}
The color difference errors, $\Delta x$ and $\Delta y$, were
defined in equation (\ref{eq:colordif_er}) and the text that follows
it. Figure \ref{fig:red} shows the color selection regions defined by
(C1) and (C2). In particular, Figure \ref{fig:red} indicates
that the condition (C2) can identify lens systems with very
large differential reddening, $\Delta E(B-V)\sim 0.3$. The cut at
$D(u-g)=0.2+\Delta D(u-g)$ is added to reduce the contamination of
quasar-star pairs. We select quasars that satisfy one or both the
criteria (C1) or (C2) as lens candidates. 

\section{Testing Selection Algorithm}
\label{sec:sim}

\begin{figure*}
\epsscale{0.8}
\plotone{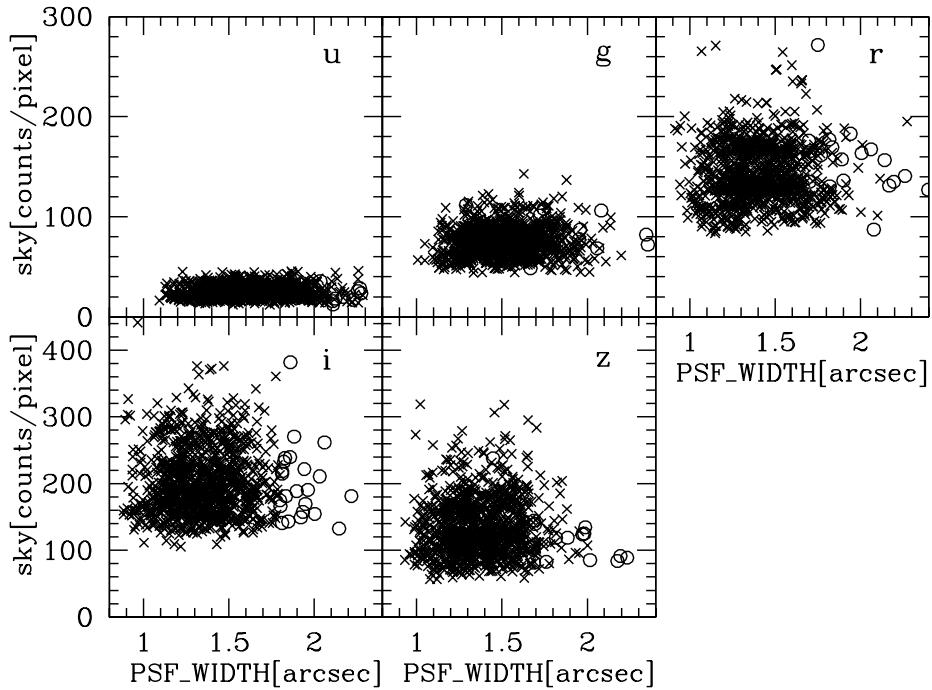}
\caption{The width of PDFs versus sky levels for 1000 randomly selected
fields in the SDSS for each band. Open circles indicate those fields
that are rejected because of poor seeing.
\label{fig:field}}
\end{figure*}

\subsection{Simulation Method}
\label{sec:sim_method}

To quantify the completeness of our selection algorithm, we perform a
series of simulations that mimic SDSS observations. We follow the
methodology developed by P03, but in this paper we implement
multicolor ($ugriz$) simulations to test our selection based both on
the morphology and color match.

First, we prepare object fields in which simulated lensed images are
placed. We assign the characteristics of each field (sky level,
seeing, and magnitude zero-point) from a randomly selected field in the
real SDSS data. Figure \ref{fig:field} shows the sky counts and seeing
sizes ({\tt PSF\_WIDTH}) of 1000 random fields. Following the
discussion in \S  \ref{sec:source}, we reject fields with {\tt
PSF\_WIDTH} $>1\farcs8$ in $i$-band in our simulations. 

\begin{figure}
\epsscale{0.98}
\plotone{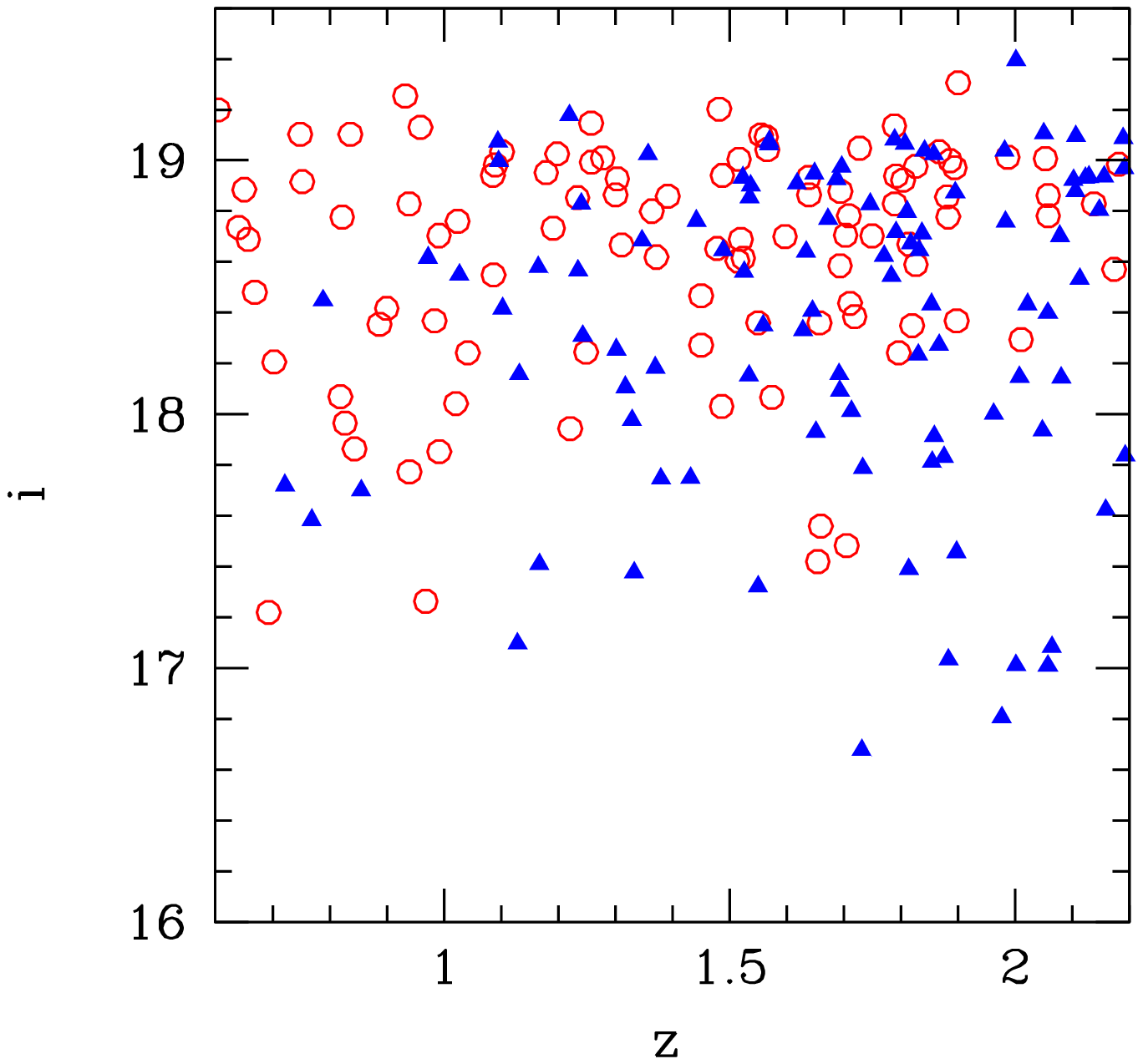}
\caption{Redshifts versus $i$-band magnitudes of randomly selected
quasars at $0.6\leq z\leq 2.2$ and $15.0\leq i_{\rm cor}\leq 19.1$ are
plotted. Open circles are selected from the quasar sample without any
weight, and filled triangles used a lensing probability weight.
\label{fig:qso}}
\end{figure}

For each field, we also choose a quasar randomly from the subset of
the DR3 quasar catalog with $0.6\leq z\leq 2.2$ and $15.0\leq i_{\rm
cor}\leq 19.1$, and adopt its redshift and magnitudes in the simulation. 
However, the redshift and magnitude distributions of the lensed
population are quite different from those of the original unlensed
population, because the lensing probability is a strong function 
of redshift and magnitude \citep[e.g.,][]{turner84}. We account
for this difference by assigning to each object a weight proportional
to its lensing probability, computed from a singular isothermal sphere
model for the mass distribution of lensing galaxies and the velocity
function of early-type galaxies derived from the SDSS
\citep{sheth03,mitchell05}. The velocity function in comoving units is
assumed to be constant with redshift. The magnification bias is
computed by assuming the luminosity function of quasars obtained from
the combination of the Two-Degree Field and SDSS results
\citep[][``2SLAQ+Croom et al.'' model]{richards05}. Although the
lensing probability is sensitive to cosmological parameters, their
main effect is to change the overall amplitude of the lensing 
probability and to have little effect on the dependence of lensing
probability on the source redshift and magnitude, on which we are
focusing. Figure \ref{fig:qso} plots redshifts versus $i$-band
magnitudes for randomly selected quasars with and without lensing
probability weighting. Because of magnification bias, gravitational
lensing preferentially chooses brighter and higher-redshift quasars.

We convert magnitude to the number of counts with the following
equations \citep{lupton99}:   
\begin{equation}
m=-\frac{2.5}{\ln(10)}\left[{\rm
    asinh}\left(\frac{F/F_0}{2b}\right)+\ln(b)\right],
\label{eq:mag1}
\end{equation}
\begin{equation}
\frac{F}{F_0}=\frac{{\rm counts}}{{\rm exptime[sec]}}10^{0.4(Z+kX)},
\label{eq:mag2}
\end{equation}
where $b$ is the softening parameter, $Z$ is the magnitude zero-point,
$k$ is the extinction coefficient, $X$ is airmass, and the
exposure time is 53.91 sec. Gravitationally lensed images are
created in the simulated image by placing two point-source objects
that are separated by an angle  $\theta$ and have a flux ratio $f_i$. 
For the point-spread function (PSF), we use an analytic PSF composed
of two Moffat functions with $\beta_1=7$ and $\beta_2=2$ and widths
given for each field (\citealt{racine96}; P03).\footnote{The width of
 the PSF measured by PHOTO ({\tt PSF\_WIDTH}) is systematically larger
 than than the input FWHM of the Moffat function by a factor $1.15$
 (P03). We take this difference into account throughout the paper.}
The PSF magnitude of each system is normalized to that of the input
quasar PSF magnitude in each band.  With this procedure, we can
simulate lensed quasars with realistic colors and field conditions.  

\begin{figure*}
\epsscale{0.7}
\plotone{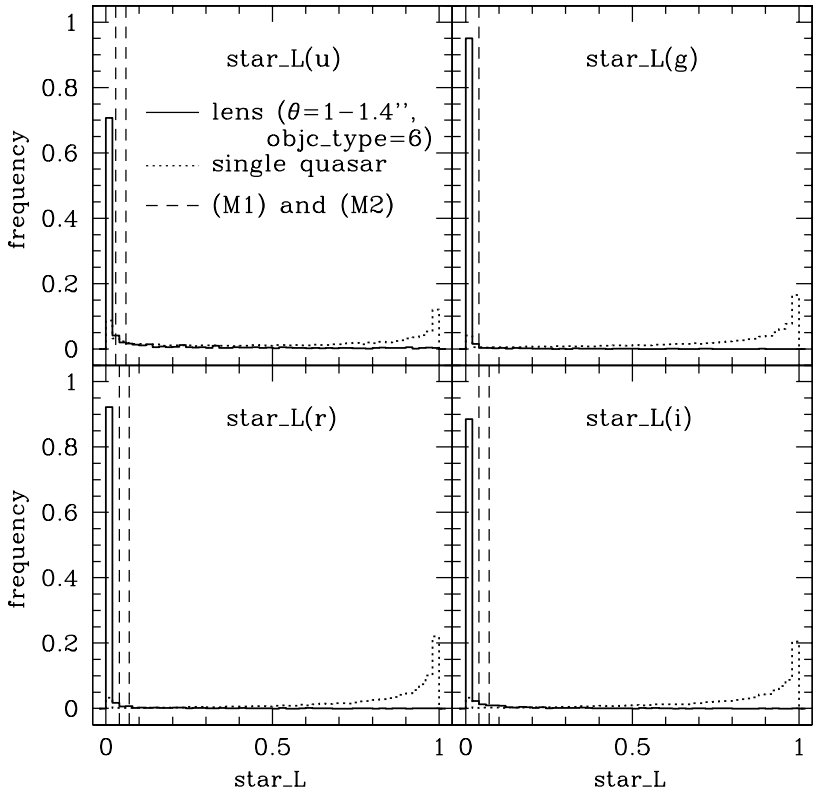}
\caption{The distributions of star likelihood {\tt star\_L} for lensed
 and single quasars, estimated from our simulations. For lensed quasars,
 we consider only those with {\tt objc\_type} $=6$, image separations
 $\theta=1''-1.4''$, and flux ratios $f_i>10^{-0.5}$. Vertical dashed
 lines indicate the limits used in our morphological selection (M1) and
 (M2) (see \S \ref{sec:sel_mor}).
\label{fig:hist_lstar}}
\end{figure*}

We also include differential reddening in our simulation. The value of
differential extinction $\Delta E(B-V)$ is randomly distributed as a
Gaussian with a standard deviation of $0.1$ mag
\citep{falco99}. Again, the reddening vectors are computed from the
composite quasar spectrum of \citet{vandenberk01} and the extinction
curve of \citet{cardelli89} assuming $R_V=3.1$. The lens redshifts are
also assigned randomly in the range $[0.2,z_{\rm   max}]$ with
$z_{\rm max}={\rm min}(0.8,z_s-0.2)$, where $z_s$ is the redshift of
the source quasar. The distribution has the average redshift $z=0.5$,
which is consistent with the value $z\sim 0.51$, the mean of 30
existing gravitationally lensed quasar systems with known lens
redshifts. We note the flux ratio $f_i$  depends on the band when
reddening is included; hereafter we refer to $f_i$ as the flux ratio of
the $i$-band image.  
  
We next pass these simulated fields through the PHOTO pipeline. Here
we briefly summarize the procedure; for more details, see P03. We
begin by preparing a PSF frame in which bright PSF stars (constructed
from two Moffat functions) are displayed. This PSF field is passed
through the fake stamp pipeline and postage stamp pipeline to generate
a psField file in which the result of the PSF fit is stored.  The
object field is passed through the fake stamp pipeline and frames
pipeline; this produces tables of the measured object parameters
(fpObjc) as well as the cropped atlas images (fpAtlas). We do not need
to pass the result through the photometric calibration pipeline, since
we can convert counts to magnitudes again by using equations
(\ref{eq:mag1}) and (\ref{eq:mag2}). The magnitude error is estimated
from that of the counts as 
\begin{equation}
\Delta m=-\frac{2.5}{\ln(10)}\frac{\Delta({\rm counts})}{{\rm
exptime[sec]}}\frac{1}{2b}\frac{10^{0.4(Z+kX)}}
{\sqrt{1+\left[(F/F_0)/2b\right]^2}}.
\label{eq:mag3}
\end{equation}
The image parameters used to select the lens candidates are extracted
from the fpObjc file. When PHOTO deblends two images, we regard the
brighter image as targeted by the quasar spectroscopic target
selection, and use the image parameters of the brighter image.

We repeat the simulation for different sets of field parameters,
redshifts and magnitudes of quasars, and reddening, and compute the
completeness of our selection algorithm as a function of the image
separation $\theta$ and flux ratio $f_i$. Each simulation run consists
of simulations of $25\times21$ quasars; the separation is changed from
$0''$ to $4\farcs 8$ with a step of $0\farcs2$ and the flux ratio from 
$10^{-1}$ to $1$ with a step of a factor of $10^{0.05}$. Since the
maximum image separation $\theta=4\farcs8$ is large enough compare with
the seeing sizes of the SDSS images, we think the completenesses of
larger-separation lenses can be approximated by those of
$\theta=4\farcs8$. We note that only a few known lensed quasars have the
flux ratios $f_i<0.1$. We simulate 1000 runs to make an accurate
estimation of the completeness.  

We note that our simulations differ from those of P03 in several
ways. In P03, the selection algorithm was tested for a few choices of
seeing sizes and magnitudes of quasars, while in this paper we
distributed them in a realistic way. We also simulated images in all
five bands, while P03 did not consider differences between different bands.

\begin{figure}
\epsscale{1.0}
\plotone{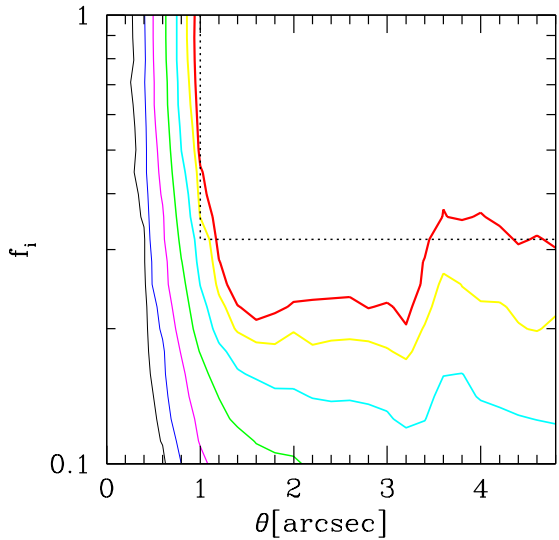}
\caption{The completeness of our lens selection algorithm in the
  image separation $\theta$-flux ratio $f_i$ plane, computed from the
  simulation described in \S \ref{sec:sim_method}. From thick to thin
  lines, we plot completeness contours of $0.95$, $0.9$, $0.75$, $0.5$,
  $0.25$, $0.1$, and $0.05$. The dotted lines indicates the limit
  above which the selection is almost complete.
\label{fig:cont}}
\end{figure}

\subsection{Star Likelihood}
\label{sec:starl}

Before showing the completeness, we check the distribution of {\tt
star\_L} to see the validity of using this parameter to identify lens
candidates.  All single quasars should be classified as point
sources ({\tt objc\_type} $=6$), thus we can choose quite
loose conditions for selecting lens candidates, without a significant
decrease of efficiency, when they are classified as extended objects 
({\tt objc\_type} $=3$; see (M3) in \S \ref{sec:sel_mor}). Therefore,
in this section we concentrate our attention on lenses with {\tt
 objc\_type} $=6$, for which we need to adopt a good discriminator
between lenses and single quasars.  

In Figure \ref{fig:hist_lstar}, we show the distributions of {\tt
star\_L} in $u$, $g$, $r$, and $i$ for both lensed and single quasars
from our simulations. For lensed quasars, we only consider the image
separation range of $\theta=1''-1.4''$ and the flux ratio
$f_i>10^{-0.5}$. It is clear that the distribution of {\tt star\_L}
for lensed quasars is markedly different from that for single
quasars. We find that lensed quasars have quite small star likelihoods
{\tt star\_L} $\sim 0$ even when lenses are classified as point
sources. On the other hand,  the distribution of star likelihoods for
single quasars peaks at {\tt star\_L} $\sim 1$ and rapidly decreases
as {\tt star\_L} decreases. This Figure assures that the conditions
(M1) and (M2) presented in \S  \ref{sec:sel_mor} serve as an excellent
method to distinguish gravitational lenses from normal quasars.  

\subsection{Completeness}
\label{sec:compl}

We show the completeness in the $\theta$-$f_i$ plane in Figure
\ref{fig:cont}. As expected, lenses with larger image separations and 
flux ratios closer to unity are more likely to be selected. In
particular, our selection algorithm is $\gtrsim 95\%$ complete at
$\theta\geq 1''$ and $f_i\geq 10^{-0.5}$ (indicated by dotted
lines). Therefore, we propose to construct a statistical sample within
this region of parameter space.   

\begin{figure*}
\epsscale{0.8}
\plotone{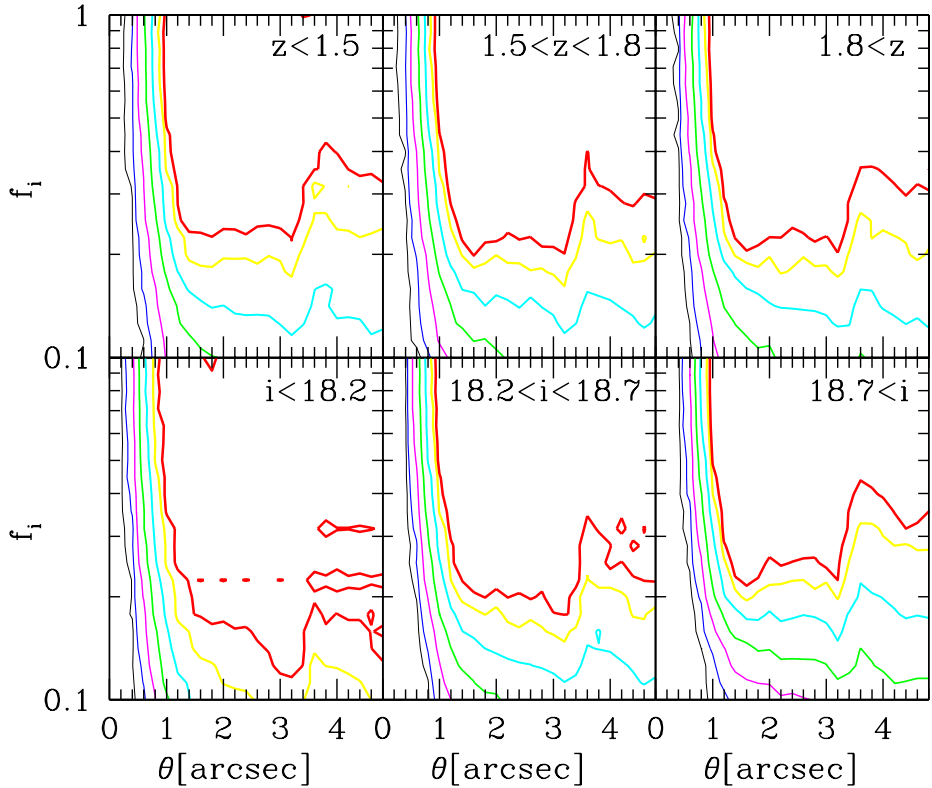}
\caption{The completeness of our lens selection algorithm for different 
redshift and magnitude bins. The definition of lines is same as Figure
  \ref{fig:cont}. {\it Upper panels:} The completenesses for restricted
 redshift ranges of $z<1.5$ ({\it left}), $1.5<z<1.8$ ({\it center}),
 and $1.8<z$ ({\it right}). {\it Lower panels:}  The completenesses for 
 restricted $i$-band magnitude ranges of $i<18.2$ ({\it left}),
 $18.2<i<18.7$ ({\it center}), and $18.7<i$ ({\it right}). In both
 cases, the sizes of bins are determined such that the (lensing
 probability weighted) number of quasars in each bin becomes roughly
 equal. 
\label{fig:cont_zm}}
\end{figure*}

Figure \ref{fig:cont_zm} shows the completenesses for three different
ranges of redshifts/magnitudes. We find that the completeness is almost
independent of the quasar redshift. On the other hand, the
completeness shows a weak but significant trend with magnitude, in the
sense that brighter quasars are more complete.  In particular,
the completeness at small $f_i$ is larger for brighter quasars. This
is reasonable since companion pairs, which are often hidden when
$f_i$ is small, are more easily detected for 
brighter quasars. However, even for faintest quasars ($i>18.7$) the
completeness at $\theta\geq 1''$ and $f_i\geq10^{-0.5}$ is very high.  
Therefore we conclude that our selection does not affect the
distributions of redshifts and magnitudes of lensed quasars very much. 

\begin{figure*}
\epsscale{0.8}
\plotone{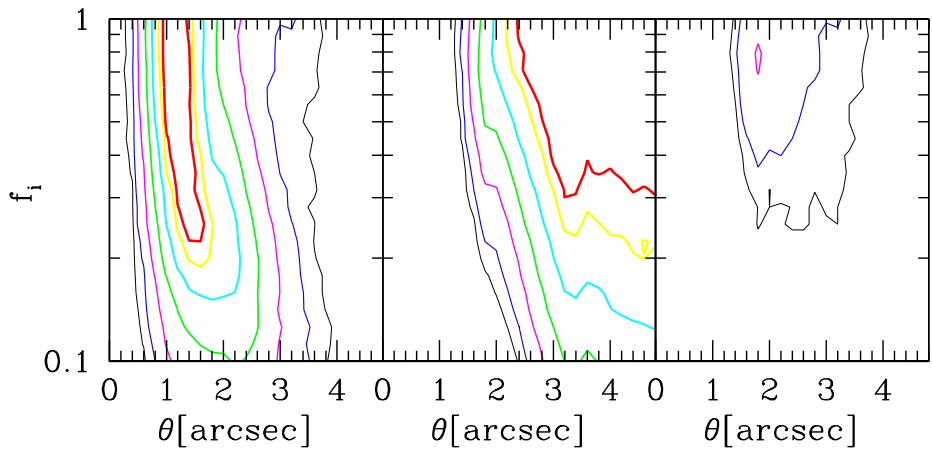}
\caption{Left: The completeness of the morphological selection
  only. Center: The completeness of the color selection only.
  Right: The fraction of lenses selected from {\it both} morphological
  and color selections. The definition of lines is same as Figure
  \ref{fig:cont}. 
\label{fig:cont_dec}}
\end{figure*}

\begin{deluxetable}{cccccc}
\tablewidth{0pt}
\tablecaption{Completeness for double lenses\label{table:comp_doub}}  
\tablehead{\colhead{$\log(f_i)$} & \colhead{$\theta$} &
 \multicolumn{4}{c}{Completeness} \\ \cline{3-6}
\colhead{} & \colhead{[arcsec]} 
& \colhead{total} & \colhead{morph} & \colhead{color}& \colhead{both}} 
\startdata 
$-1.00$ & 0.0 & 0.036 & 0.036 & 0.000 & 0.000 \\
$-1.00$ & 0.2 & 0.031 & 0.031 & 0.000 & 0.000  \\
$-1.00$ & 0.4 & 0.032 & 0.032 & 0.000 & 0.000  \\
$-1.00$ & 0.6 & 0.041 & 0.041 & 0.000 & 0.000  \\
$-1.00$ & 0.8 & 0.100 & 0.100 & 0.000 & 0.000  \\
$-1.00$ & 1.0 & 0.213 & 0.213 & 0.000 & 0.000  \\
$-1.00$ & 1.2 & 0.304 & 0.304 & 0.000 & 0.000  \\
$-1.00$ & 1.4 & 0.378 & 0.378 & 0.000 & 0.000  \\
$-1.00$ & 1.6 & 0.415 & 0.415 & 0.001 & 0.001  \\
$-1.00$ & 1.8 & 0.461 & 0.459 & 0.005 & 0.003  \\
$-1.00$ & 2.0 & 0.472 & 0.463 & 0.011 & 0.002  \\
\enddata
\tablecomments{Full table to appear in the on-line edition.}
\end{deluxetable}

Figure \ref{fig:cont} does not make it clear if lenses are mainly
chosen by the morphological or color selection at each point in the
$\theta$-$f_i$ plane. Figure \ref{fig:cont_dec} shows the
completeness from the two selection algorithms separately. We also show
the overlap  of these two selections. At $\theta\lesssim 2''$ lenses are
mostly selected morphologically, and at $\theta\gtrsim 2''$ the color
selection dominates. The overlap of these two selections is quite
small. The completeness from the two selection algorithms as well as
total completeness are listed in Table \ref{table:comp_doub}.

\subsection{Magnification Factor}

The magnification bias is one of the most important elements in
predicting lensing probabilities. The SDSS quasar target selection
\citep{richards02} uses PSF magnitudes for the magnitude limit,
therefore we examine those of simulated lensed quasars with various
image separations. Basically, when the image separation is quite large
and two images are deblended successfully by PHOTO, the PSF magnitude
of the targeted quasar should be roughly equal to that of the brighter
image. However, when the image separation is too small to be
distinguished from a point source, the PSF magnitude is that of the
sum of the two images. Bearing this in mind, we define the following
parameter:  
\begin{equation}
\bar{\mu}\equiv\frac{\mu-\mu_{\rm bri}}{\mu_{\rm fai}},
\label{eq:mbias}
\end{equation}
where $\mu_{\rm bri}$ and $\mu_{\rm fai}$ are the magnification factors
for the brighter and fainter images, respectively. The magnification
factor $\mu$ indicates the one measured in the SDSS (i.e., the PSF
magnitude of the lens system divided by the original magnitude of the
source quasar). The definition indicates that $\mu=\mu_{\rm
  tot}\equiv\mu_{\rm bri}+\mu_{\rm fai}$ and $\mu=\mu_{\rm bri}$
correspond to $\bar{\mu}=1$ and $\bar{\mu}=0$,
respectively. Therefore, the parameter should approach zero at large
separations, and $\bar{\mu}\rightarrow 1$ for $\theta \rightarrow0$.  

\begin{figure}
\epsscale{1.0}
\plotone{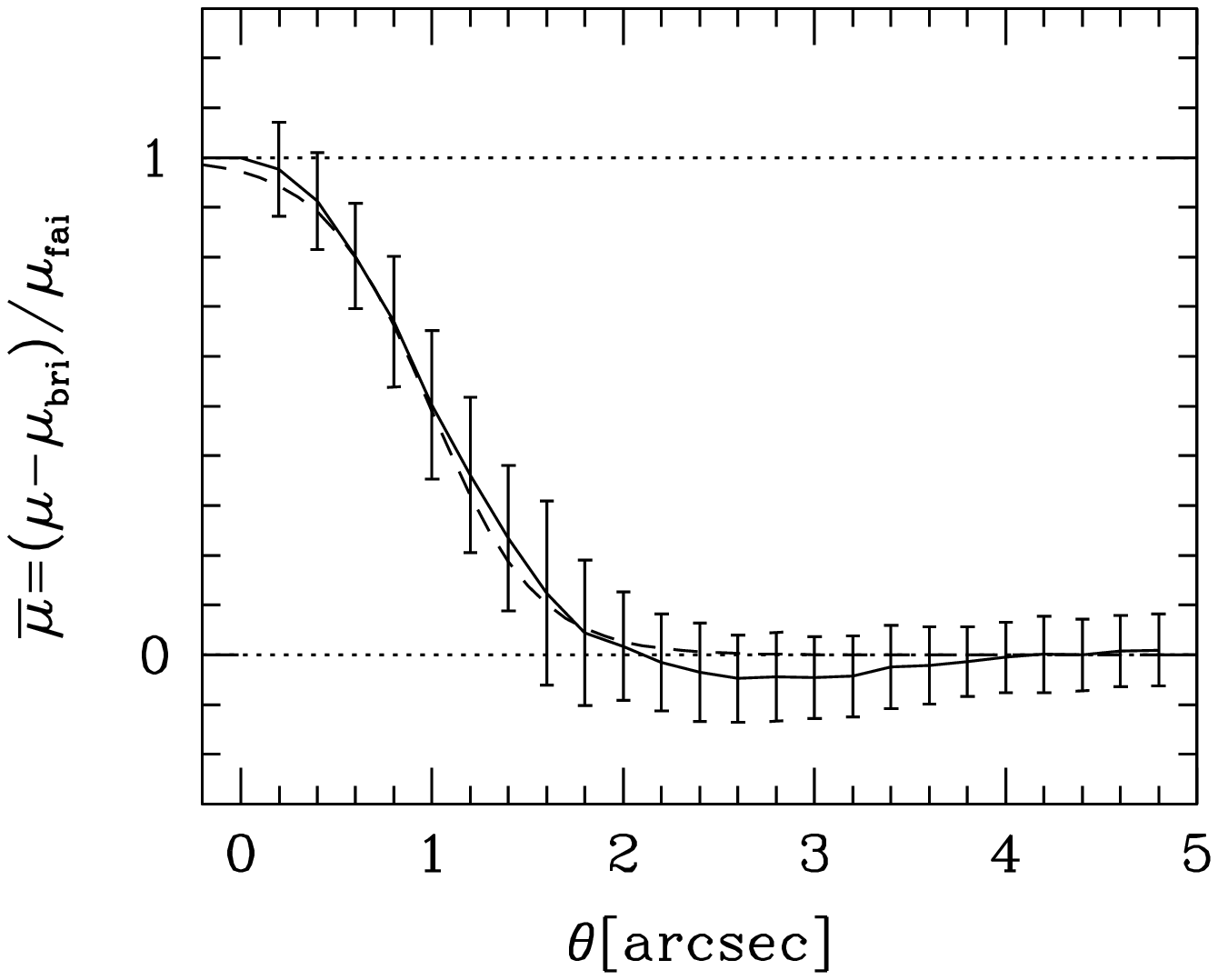}
\caption{The magnification factor of lensed quasars. The parameter
  $\bar{\mu}$ (eq. [\ref{eq:mbias}]) is plotted as a function of
  image separation $\theta$. For each image separation the mean and
  its standard deviation are shown. The dashed line indicates the
  analytic fit (eq. [\ref{eq:mbias_fit}]).
\label{fig:flux}}
\end{figure}

We derive $\bar{\mu}$ as a function of the image separation 
from our simulation runs. We only consider simulated lens systems with
flux ratio $f_i\geq10^{-0.5}$, because the derived  $\bar{\mu}$
contains large errors when flux ratios are very large $\mu_{\rm
  bri}\gg\mu_{\rm fai}$, and also because the completeness is low for
large flux ratio systems. Moreover, $\mu_{\rm tot}\sim \mu_{\rm bri}$
at $f_i\ll1$ implies that the magnification factors do not change as a
function of the image separation in any case. Figure \ref{fig:flux}
plots the result. As expected, $\bar{\mu}$ changes from $\bar{\mu}\sim
1$ to $\sim 0$ with increasing image separations. We find that the
curve is fitted roughly by the following form: 
\begin{equation}
\bar{\mu}=\frac{1}{2}\left[1+\tanh\left(1.76-1.78\theta\right)\right],
\label{eq:mbias_fit}
\end{equation}
where the image separation $\theta$ is in units of arcsec. 

We note that for a given image separation the magnification bias can be
written as
\begin{equation}
B=\bar{\mu}B_{\rm tot}+(1-\bar{\mu})B_{\rm bri},
\label{eq:bias}
\end{equation}
where $B_{\rm tot}$ and $B_{\rm bri}$ are the magnification bias from the
total magnification and magnification of the brighter image, respectively.

\subsection{Quadruple Lens Systems}

\begin{deluxetable*}{ccccc}
\tablewidth{0pt}
\tablecaption{Quadruply lensed quasars used in the simulation\label{table:quad}}  
\tablehead{\colhead{Name} & \colhead{Type\tablenotemark{a}} &
  \colhead{$\theta$\tablenotemark{b}} & 
  \colhead{$f_{\rm quad}$\tablenotemark{c}} & \colhead{Reference\tablenotemark{d}}}
\startdata 
PG 1115+080      & fold & $2\farcs43$ & 0.17 & CASTLES\tablenotemark{e} \\
SDSS J0924+0129  & fold & $1\farcs81$ & 0.47 & \citet{keeton06} \\
SDSS J1004+4112  & fold & $14\farcs6$ & 0.23 & \citet{oguri04a} \\
WFI2026$-$4536  & fold & $1\farcs44$ & 0.31 & \citet{morgan04} \\
WFI2033$-$4723  & fold & $2\farcs53$ & 0.56 & \citet{morgan04} \\
HE0230$-$2130   & fold & $2\farcs19$ & 0.59 & CASTLES\tablenotemark{e} \\
MG0414+0534     & fold & $2\farcs13$ & 0.47 & CASTLES\tablenotemark{e} \\
B0712+472       & fold & $1\farcs29$ & 0.20 & CASTLES\tablenotemark{e} \\
B1555+375       & fold & $0\farcs42$ & 0.49 & \citet{marlow99} \\
B1608+656       & fold & $2\farcs10$ & 0.49 & \citet{fassnacht02} \\
H1413+117       & cross& $1\farcs35$ & 0.71 & CASTLES\tablenotemark{e} \\
HE0435$-$1223   & cross& $2\farcs56$ & 0.62 & \citet{wisotzki02} \\
HST12531$-$2914 & cross& $1\farcs38$ & 0.92 & CASTLES\tablenotemark{e} \\
HST14113+5211   & cross& $2\farcs26$ & 0.82 & CASTLES\tablenotemark{e} \\
HST14176+5226   & cross& $3\farcs26$ & 0.79 & CASTLES\tablenotemark{e} \\
Q2237+030       & cross& $1\farcs83$ & 0.31 & CASTLES\tablenotemark{e} \\
RX J0911+0551    & cusp & $3\farcs25$ & 0.31 & CASTLES\tablenotemark{e} \\
RX J1131$-$1231  & cusp & $3\farcs24$ & 0.09 & \citet{sluse03} \\
B0128+437       & cusp & $0\farcs55$ & 0.53 & \citet{{phillips00}} \\
B1422+231       & cusp & $1\farcs29$ & 0.03 & CASTLES\tablenotemark{e} \\
\enddata
\tablenotetext{a}{Based on visual inspection.} 
\tablenotetext{b}{The maximum separation between multiple images.} 
\tablenotetext{c}{The flux ratio between the brightest image and the
 image farthest from the brightest image.} 
\tablenotetext{d}{The reference from which we adopt the image
 positions and fluxes for the simulation. For fluxes, we use those of
 the $i$-band (F814W for the Hubble Space Telescope data) image when the
 optical data are available. If not, we adopt those of the radio band image.} 
\tablenotetext{e}{C.~S. Kochanek, E.~E. Falco, C. Impey, J. Lehar,
 B. McLeod, H.-W. Rix, CASTLES survey, http://cfa-www.harvard.edu/castles/} 
\end{deluxetable*}

Thus far we have concentrated our attention on two-component lenses,
but in reality a non-negligible fraction of gravitationally lensed
quasars have four components. In this subsection we consider our
selection algorithm's effectiveness for such quadruple lenses. 

The difficulty in studying the selection of quadruple lenses lies in
their complexity; while double lenses are parameterized by flux ratio
and image separation, we require 8 parameters to characterize a
quadruple lens, making a full exploration almost impossible.
Instead, in this paper we adopt the image configurations and flux
ratios from 20 known quadruple lenses taken from the CASTLES webpage (see
Table \ref{table:quad}). We apply a similarity transformation to the
quadruple lenses in order to derive the  completeness as a function of
the image separation. We define the image separation to be the maximum
distance between multiple images. The differential reddening is
included, again using the distribution of \citet{falco99}. 
As in the case of double lenses, we simulate 1000 runs to make an
accurate estimate of the completeness. 

\begin{figure}
\epsscale{1.0}
\plotone{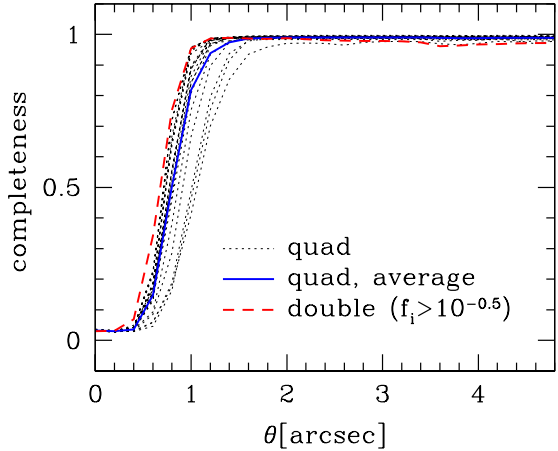}
\caption{The completeness for quadruple lenses as a function of
 the image separation. The dotted line shows the completeness of each
 quadruple lens image configuration and flux ratios taken from real
 observational data (see Table \ref{table:quad}). The thick solid line
 indicates the completeness averaged over the 20 dotted lines. The
 completeness of double lenses, averaged over the flux ratios in the
 range $10^{-0.5}<f_i<1$, is also shown for reference.
\label{fig:quad}}
\end{figure}

Figure \ref{fig:quad} plots the completeness of each of
the 20 quadruple lens configurations as a function of the image
separation, as well as the average of them. We find that our algorithm
can select quadruple lenses reasonably well: Even in the limiting
case, $\theta=1''$, the completeness is $82\%$ on average, and at
$\theta>2''$ it is almost unity. We note that in both cases the
completenesses vary most rapidly at $\theta\sim 1''$ where the number of
lensed quasars peaks. There is a systematic difference of
the completeness between double and quadruple lenses: At $\theta<2''$ the
completeness for double lenses is larger than that for quadruple lenses,
while at $\theta>2''$ the quadruple lenses are more complete. The
differences are small, but they might become important when discussing
the relative number of quadruple and double lenses
\citep{keeton97,rusin01,chae03,oguri04c}.  

\begin{figure*}
\epsscale{0.8}
\plotone{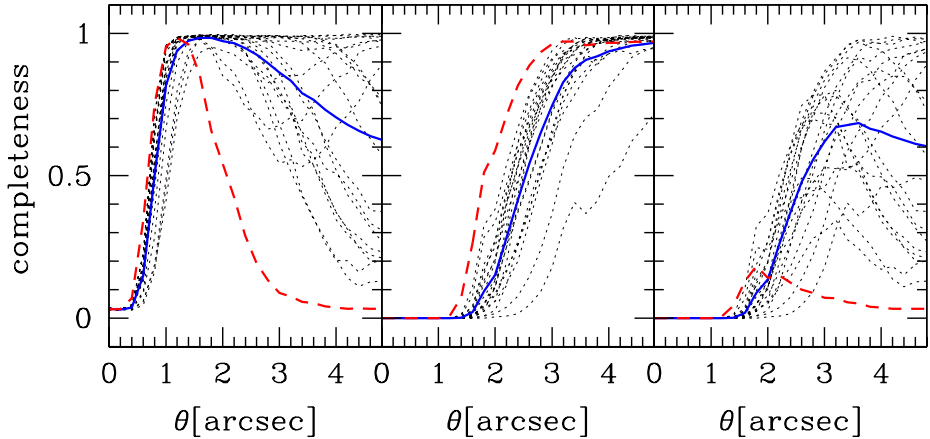}
\caption{The completeness for quadruple lenses, with the morphological
 selection only ({\it left}), the color selection only ({\it center}),
 and {\it both} morphological and color selections ({\it right}). The
 definition of lines is same as in Figure \ref{fig:quad}.  
\label{fig:quad_dec}}
\end{figure*}

Figure \ref{fig:quad_dec} shows the completeness for morphological and
color selections separately. Unlike double lenses, quadruple lenses
are effectively chosen by the morphological selection even at large
image separations, $\theta>3''$. In addition, it is quite common that
lenses are selected by both criteria This is because quadruple lenses
are more complex than double lenses. In particular, the complexity and
larger number of images for quadruple lenses make the completeness at
$\theta>2''$ quite large. Again, the completeness from the two
selection algorithms as well as total completeness are listed in Table
\ref{table:comp_quad}.  

\begin{figure}
\epsscale{1.0}
\plotone{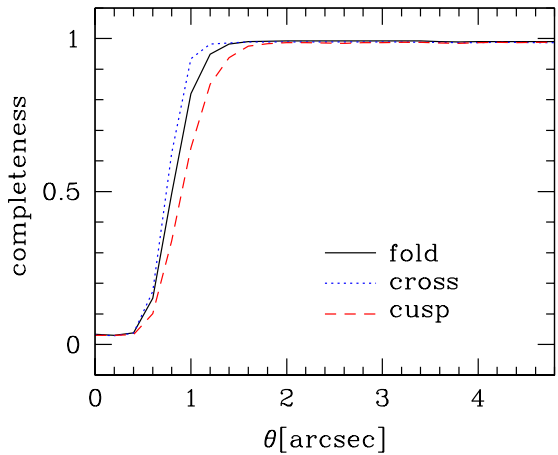}
\caption{The completeness for quadruple lenses of different image
 configuration types: fold ({\it solid}) , cross ({\it dotted}), and
 cusp ({\it dashed}).
\label{fig:quad_type}}
\end{figure}

\begin{figure}
\epsscale{1.0}
\plotone{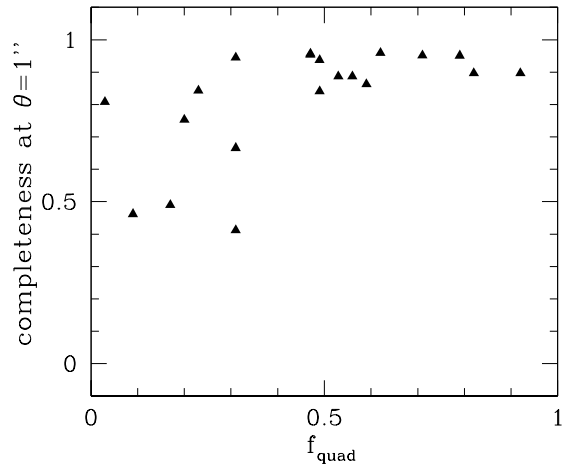}
\caption{The completeness at $\theta=1''$ and $f_{\rm quad}$ (the flux
 ratio between the brightest image and the image farthest from the
 brightest image) for 20 known quadruple lenses. The values of $f_{\rm
 quad}$ are listed in Table \ref{table:quad}.
\label{fig:quad_selflu}}
\end{figure}

We also study the completeness for different image configuration types:
fold (two of the images lie close together), cross (almost symmetric),
and cusp (three of the images lie close together). To do so, we divide
the 20 quadruple lenses into these three categories by visual inspection,
and compute the average completeness for each category.  The
categorization is summarized in Table \ref{table:quad}, and the
results are shown in Figure \ref{fig:quad_type}. Our algorithm selects
cross lenses very well, but is less effective for cusp lenses; this is
because the fourth image of a cusp lens, which lies far from the
critical curve, is usually much fainter than the other images near the
critical curve and therefore the effective ``flux ratio'' of cusp
lenses becomes small. To pursue this further, we consider the flux ratio
between the brightest image and the image farthest from the brightest
image (denoted by $f_{\rm quad}$), and see its correlation with the
completeness. Figure \ref{fig:quad_selflu} plots $f_{\rm quad}$ and
the completeness at $\theta=1''$ for all 20 quadruple lenses listed in
Table \ref{table:quad}. As expected, we see a clear correlation: All
quadruple lenses with completeness $<0.8$ have $f_{\rm
  quad}<0.4$. Therefore, despite the complexity and large number of
parameters involved in quadruple lenses, $f_{\rm quad}$ gives a rough
criterion of how well our selection algorithm can identify quadruple
lenses. 

\begin{deluxetable}{cccccc}
\tablewidth{0pt}
\tablecaption{Completeness for quadruple lenses\label{table:comp_quad}}  
\tablehead{\colhead{Name} & \colhead{$\theta$} &
 \multicolumn{4}{c}{Completeness} \\ \cline{3-6}
\colhead{} & \colhead{[arcsec]} 
& \colhead{total} & \colhead{morph} & \colhead{color}& \colhead{both}} 
\startdata 
PG1115+080 & 0.0 & 0.035 & 0.035 & 0.000 & 0.000  \\
PG1115+080 & 0.2 & 0.030 & 0.030 & 0.000 & 0.000  \\
PG1115+080 & 0.4 & 0.029 & 0.029 & 0.000 & 0.000  \\
PG1115+080 & 0.6 & 0.059 & 0.059 & 0.000 & 0.000  \\
PG1115+080 & 0.8 & 0.188 & 0.188 & 0.000 & 0.000  \\
PG1115+080 & 1.0 & 0.490 & 0.490 & 0.000 & 0.000  \\
PG1115+080 & 1.2 & 0.796 & 0.796 & 0.000 & 0.000  \\
PG1115+080 & 1.4 & 0.937 & 0.937 & 0.000 & 0.000  \\
PG1115+080 & 1.6 & 0.978 & 0.978 & 0.001 & 0.001  \\
PG1115+080 & 1.8 & 0.987 & 0.987 & 0.007 & 0.007  \\
PG1115+080 & 2.0 & 0.992 & 0.991 & 0.016 & 0.015  \\
\enddata
\tablecomments{Full table to appear in the on-line edition.}
\end{deluxetable}

\section{Removing Single Quasars}
\label{sec:second}

\begin{figure*}
\epsscale{0.75}
\plotone{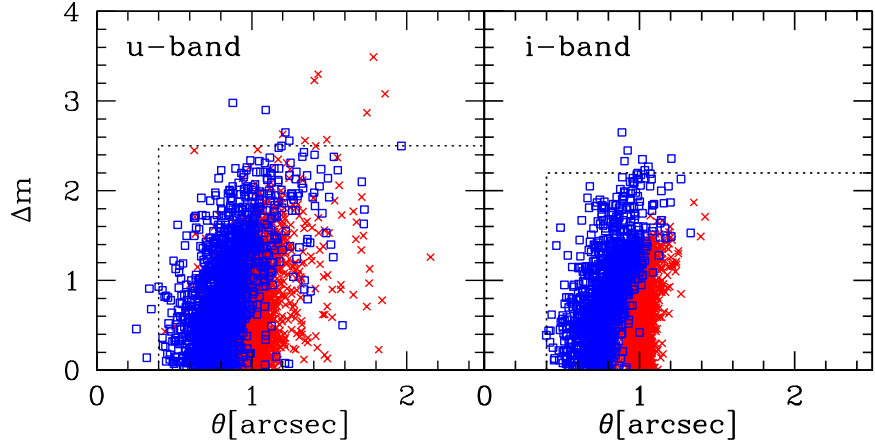}
\caption{The distribution of inferred magnification differences
 ($\Delta m$) and image separation ($\theta$) of simulated images,
 derived from fitting by two PSFs with GALFIT software \citep{peng02}.  
  The true image separation and the $i$-band flux ratio are set to
 $\theta=1''$ and $f_i=[10^{-0.5},1]$ (for double lenses), respectively. 
 Crosses indicate double lenses, and open squares are quadruple
 lenses. The left and right panels show the results of $u$-band and
 $i$-band images, respectively. The dotted lines indicate our proposed
 limit in the image separation-flux ratio space (S1) defined in
 equation (\ref{secsel}). 
 \label{fig:sepsim}} 
\end{figure*}

\begin{figure}
\epsscale{1.0}
\plotone{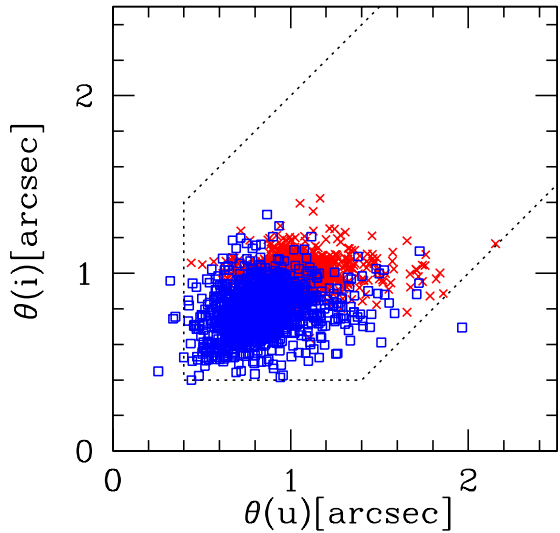}
\caption{The distribution of $u$- and $i$-band image separations
 ($\theta$) of simulated images with $\theta=1''$ and  the $i$-band flux
 ratio $f_i=[10^{-0.5},1]$ (for double lenses), derived from fitting by
 two PSFs with GALFIT software \citep{peng02}. Crosses indicate double
 lenses, and open squares are quadruple lenses. The dotted lines indicate
 our proposed limit in the image separation space (S1) defined in
 equation (\ref{secsel}). 
 \label{fig:sepsim_sep}} 
\end{figure}

The nonzero completeness at $\theta\rightarrow 0$ implies that our
morphological selection algorithm falsely identifies ``normal'' quasars
as lenses. Indeed, when we apply our algorithm to DR3 quasar catalog,
we find many lens candidates, many of which appear to be point sources
(see \S \ref{sec:eff}), although most normal single quasars are
removed by our morphological selection. Therefore, in this section, we
present an additional selection which is aimed to exclude such point
sources. 

To do so, we fit systems selected by our morphological selection
algorithm with two PSFs using the GALFIT package \citep{peng02}, and
derive the image separation and flux ratio. We use stars in the same
field as a PSF template in fitting. The fitting of any single object with 
two PSFs should result in either too large magnitude differences or too
small image separations. Here we only consider gravitational lenses
included in the proposed statistical sample, i.e., those with
$\theta>1''$ and $f_i>10^{-0.5}$. The point sources should have
significantly smaller image separation and/or larger flux ratios than
the above limit. Therefore we can reject such point sources by
making a cut in image separation-flux ratio space. We do this fit on
both the $u$- and $i$-band images: We use the $i$-band image because
the flux ratio is defined by that of the $i$-band image. The $u$-band
is used because it is a powerful discriminator between lensed pairs and
superpositions of quasars and stars or quasars and galaxies. 

To determine the region of parameter space we should exclude, we
perform fitting of simulated images (see \S \ref{sec:sim_method}). We
consider only the extreme image separation, $\theta=1''$, for which
fitting by two PSFs will be most difficult. The flux ratio varied over
the range $f_i=[10^{-0.5},1]$. We show the results for both $u$- and
$i$-band simulated images in Figures \ref{fig:sepsim} (distribution in
image separation-flux ratio space for each band) and \ref{fig:sepsim_sep}
(distribution in $u$- and $i$-band image separation space). It is seen
that quadruple lenses show smaller image separations on average than
double lenses; this is clearly because GALFIT tends to fit the two
brightest images that are not necessarily the two separated farthest
from each other. From this result, we determine the selection
criterion to be the following:  
\begin{eqnarray}
\mbox{(S1):}&& \theta(u)\geq 0\farcs4, \nonumber\\
&& \theta(i)\geq 0\farcs4, \nonumber\\
&& \Delta m(u)\leq2.5, \nonumber\\
&& \Delta m(i)\leq2.2,\nonumber\\
&& \left|\theta(u)-\theta(i)\right|\leq1'',\label{secsel}
\end{eqnarray}
where $\theta$ and $\Delta m$ are the separation and magnitude
difference between fitted two PSF components. 
We select candidates that meet all of the above criteria. In our
simulation, more than 99\% of double and quadruple lenses with
$\theta=1''$ pass the condition, thus we conclude that this
additional selection hardly changes the high completeness of our
statistical lens sample estimated in \S \ref{sec:sim}. The condition
(S1) rejects $\sim 85\%$ of candidates selected by the  
morphological selection (see \S \ref{sec:eff}). 

\section{Efficiency}
\label{sec:eff}

A definitive estimate of the efficiency requires follow-up
observations of all candidates which is still in progress. Nevertheless,
in this section we discuss the efficiency of our selection algorithm 
in light of the follow-up observations that we have performed thus far.

\begin{deluxetable*}{cccccccc}
\tablewidth{0pt}
\tablecaption{Preliminary efficiency of our selection algorithm\label{table:eff}}  
\tablehead{\colhead{$\theta$} & \colhead{$N_{\rm mor}$\tablenotemark{a}} 
& \colhead{$N_{\rm mor+}$\tablenotemark{b}} 
& \colhead{$N_{\rm col}$\tablenotemark{c}} &
 \colhead{$N_{\rm cand}$\tablenotemark{d}} & 
 \colhead{$N_{\rm obs}$\tablenotemark{e}} & 
 \colhead{$N_{\rm lens}$\tablenotemark{f}} 
& \colhead{$N_{\rm lens}/N_{\rm cand}$}}
\startdata 
$<2''$     & 568 & 88 & 9   & 96  &  77 & 8  & 0.083 \\
$2''-7''$  & 74  & 8  & 59  & 66  &  62 & 2  & 0.030 \\
$7''-20''$ & 6   & 0  & 155 & 155 &  141 & 1  & 0.006 \\ \hline
$<20''$    & 648 & 96 & 223 & 317 &  280 & 11 & 0.035 \\
\enddata
\tablenotetext{a}{Number of candidates selected by the morphological
 selection.} 
\tablenotetext{b}{Number of candidates selected by the morphological
 selection plus additional selection presented in \S \ref{sec:second}.} 
\tablenotetext{c}{Number of candidates selected by the color selection.} 
\tablenotetext{d}{Total number of candidates selected by the
 morphological and color selection. It does not necessarily coincide
 with the sum of $N_{\rm mor+}$ and $N_{\rm col}$, because some of
 candidates are selected by {\it both} algorithms.}  
\tablenotetext{e}{Number of candidates for which judgement of
lens/nonlens was made on the basis of follow-up observations and/or
various survey data.} 
\tablenotetext{f}{Current number of confirmed lenses included among
  the candidates.}  
\end{deluxetable*}

To do so, we again adopt the DR3 quasar sample \citep{schneider05}. As 
discussed in \S \ref{sec:source}, we use 22,682 quasars at 
$0.6\leq z\leq 2.2$ and $15.0\leq i_{\rm cor}\leq 19.1$ as our source
quasar sample. We apply our algorithms to the quasar sample, and
select $648$ candidates from the morphological selection and
$228$ candidates from the color selection. By applying the additional
cut described in \S \ref{sec:second} to morphologically selected
candidates, the number of candidates from the morphological selection
reduces to $97$, only $\sim 15$\% of the original. For color selection, we
only selected candidates with image separations smaller than
$20\farcs1$ and magnitude differences smaller than $1.3$, since we
concentrate our attention to the lens sample in the range
$1''\leq\theta\leq 20''$ and $f_i\geq 10^{-0.5}$. It is expected that
the efficiency is a strong function of the image separation, thus we
consider the three image separation ranges, $\theta<2''$,
$2''<\theta<7''$, and $7''<\theta$, and derive the efficiency for
each image separation range separately. The image separations 
of morphologically selected candidates we used are those obtained by
fitting $i$-band SDSS images with two PSFs using GALFIT (see \S
\ref{sec:second}).  The result is summarized in
Table \ref{table:eff}. We note that the algorithm successfully
identified all known lenses included in the subset of the DR3 quasar
catalog, thereby confirming the high completeness shown in \S
\ref{sec:sim}. Since we still have candidates that are to be
observed, the efficiency shown in Table \ref{table:eff} should in
principle be regarded as the lower limit. However, we believe that
this is close to the true value because of the ``observation bias''
whereby we preferentially observe better candidates. As expected, 
the efficiency is dependent on the image separation: The efficiencies
are 8\%, 3\%, and 0.6\% for $\theta<2''$, $2''<\theta<7''$,
and $7''<\theta$, respectively. Most false positives, which have been
revealed by follow-up observations, were quasar-star pairs or quasar pairs
\citep[see also][]{hennawi06a}. For the smallest image separation bin,
the efficiency is comparable to that of the CLASS, which discovered 13
gravitational lenses out of 149 candidates \citep{browne03}. 
We note that the candidates can be restricted further by comparing
with other existing observations (see Paper II) such as
Faint Images of the Radio Sky at Twenty-cm \citep{becker95}, which
should increase the efficiency. 

\section{Conclusion}
\label{sec:con}

We have presented the SQLS lens candidate selection algorithm from the
SDSS spectroscopic quasar catalog. Our algorithm consists of 
two selection methods, a morphological selection that identifies
``extended'' quasars, and a color selection that picks up adjacent
objects with similar colors. Both selection methods rely only on imaging
parameters that the SDSS reduction pipeline generates, thereby allowing
easy and fast selection of lens candidates. To reduce the number of
candidates further, we fit the images of morphologically selected
candidates with two PSFs, deriving the image separations and magnitude
differences, and applying cuts in the image separation-magnitude
difference space. This allows us to remove most single quasars that
are otherwise selected by the morphological selection. 

We have performed simulations of lensed quasar images in the SDSS to
determine the effectiveness of the selection algorithm. In the
simulations, distributions of field parameters (i.e., seeings and sky
levels) as well as those of source quasars are taken from the real
SDSS data. We have also taken differential reddening by lens galaxies
into account. The simulated SDSS images are then passed through the
imaging data processing pipeline used in the SDSS. This allows a
reliable estimate of the completeness of the algorithm to be made. We
have found that our selection algorithm is almost complete at 
image separations larger than $1''$ and flux ratios larger than
$10^{-0.5}$. We have also quantified the 
magnification factor of lensed images as a function of the image
separation, which is important for accurate computation of magnification
bias. The algorithm successfully identifies both double and quadruple
lenses, although there is a systematic difference of completenesses
between double and quadruple lenses. The efficiency depends strongly
on the image separation; at separations less than $2''$ the preliminary
efficiency of our selection algorithm is about 8\%, comparable to that
of the CLASS.

An important effect we have neglected in our simulation is
fluxes from lens galaxies. Although for quasars with $i\lesssim 19.1$
and redshifts above $0.6$ lens galaxies tend to be fainter than lensed
quasars, they could affect  quantitative results. An important point is
that lensed quasars are not selected by the quasar spectroscopic target
selection if the fluxes of lens galaxies dominate: A good example is
the high-redshift lens SDSS J0903+5028 \citep{johnston03} that was
targeted for SDSS spectroscopy as a luminous red galaxy, rather than a
quasar, because the lensing galaxy is brighter than the lensed quasar
images. Therefore such systems will not be included in SDSS
spectroscopic quasar catalogs. This means that we need to simulate
quasar spectroscopic target selection for lensed systems, as well as
lens selection algorithms, to address the effect of lens galaxy
fluxes. This is beyond the scope of this paper. 

The well-understood selection function, together with the knowledge of 
redshift and magnitude distributions of source quasars, is very
important to control possible systematic effects. We plan to construct
a first SQLS complete lens sample from the DR3 quasar catalog, which
will be presented in Paper II. 

\begin{acknowledgments}
We thank Issha Kayo for careful reading of the manuscript. 
N.~I. is supported by JSPS through JSPS Research Fellowship for Young
Scientists. M.~A.~S. is supported by NSF grant AST 03-07409.

Funding for the SDSS and SDSS-II has been provided by the Alfred
P. Sloan Foundation, the Participating Institutions, the National
Science Foundation, the U.S. Department of Energy, the National
Aeronautics and Space Administration, the Japanese Monbukagakusho, the
Max Planck Society, and the Higher Education Funding Council for
England. The SDSS Web Site is http://www.sdss.org/. 

The SDSS is managed by the Astrophysical Research Consortium for the
Participating Institutions. The Participating Institutions are the
American Museum of Natural History, Astrophysical Institute Potsdam,
University of Basel, Cambridge University, Case Western Reserve
University, University of Chicago, Drexel University, Fermilab, the
Institute for Advanced Study, the Japan Participation Group, Johns
Hopkins University, the Joint Institute for Nuclear Astrophysics, the
Kavli Institute for Particle Astrophysics and Cosmology, the Korean
Scientist Group, the Chinese Academy of Sciences (LAMOST), Los Alamos
National Laboratory, the Max-Planck-Institute for Astronomy (MPA), the
Max-Planck-Institute for Astrophysics (MPIA), New Mexico State
University, Ohio State University, University of Pittsburgh, University
of Portsmouth, Princeton University, the United States Naval
Observatory, and the University of Washington. 
\end{acknowledgments}

\begin{appendix}

\section{Object types of lensed quasar systems}
\label{sec:append}

Gravitationally lensed quasars are sometimes classified as extended
objects ({\tt objc\_type} $=3$) by PHOTO. This is very important for
lensing of high-redshift ($z\gtrsim 3$) quasars because extended
high-redshift quasars are not targeted by the quasar target selection
pipeline (see Figure \ref{fig:compl}). Therefore, we need to take
this bias into account when we constrain the shape of the
high-redshift quasar luminosity function from the absence of strong
gravitational lensing \citep[e.g.,][]{richards06a}. In addition, this
bias strongly affects the search for lenses using a photometrically
selected quasar sample \citep{richards04}, because extended quasars
are not included in the photometric sample, again in order to enhance
the efficiency.  In this appendix, we derive the fraction of lens
systems that are classified as extended objects from our simulations
of lensed quasar images (\S \ref{sec:sim}). Although we focused our
attention on bright ($i<19.1$) low-redshift quasars and therefore the
result may not be applied to high-redshift quasars that are fainter
and have quite different colors, the result is still useful to
understand how PHOTO classifies lensed quasar images.

\begin{figure}[b]
\epsscale{1.0}
\plotone{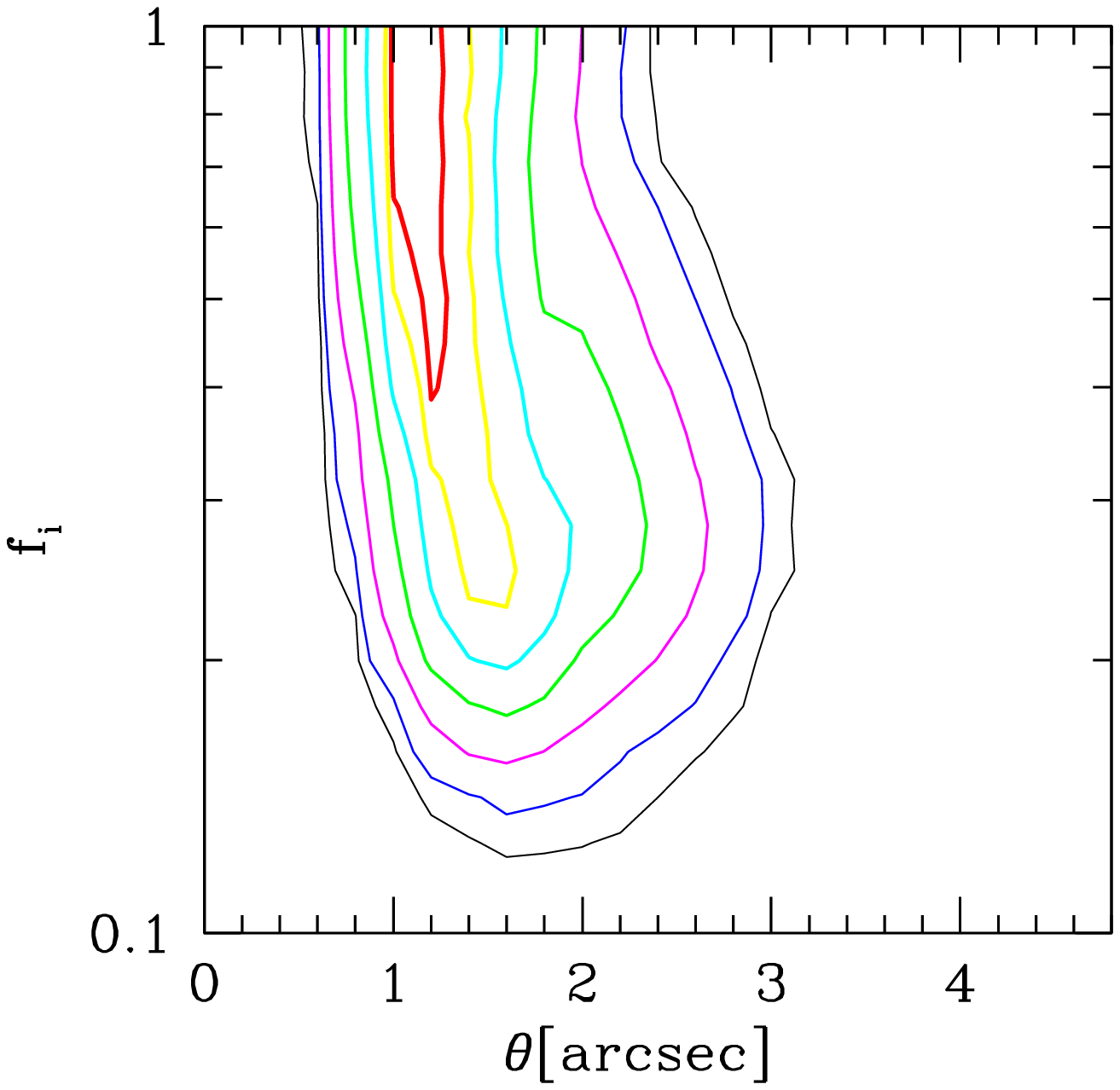}
\caption{Contours of the fraction of simulated double lenses that are
  classified as extended objects ({\tt objc\_type} $=3$). The
  definition of lines is same as in Figure \ref{fig:cont}. 
\label{fig:cont_obj}}
\end{figure}

\begin{figure}[b]
\epsscale{1.0}
\plotone{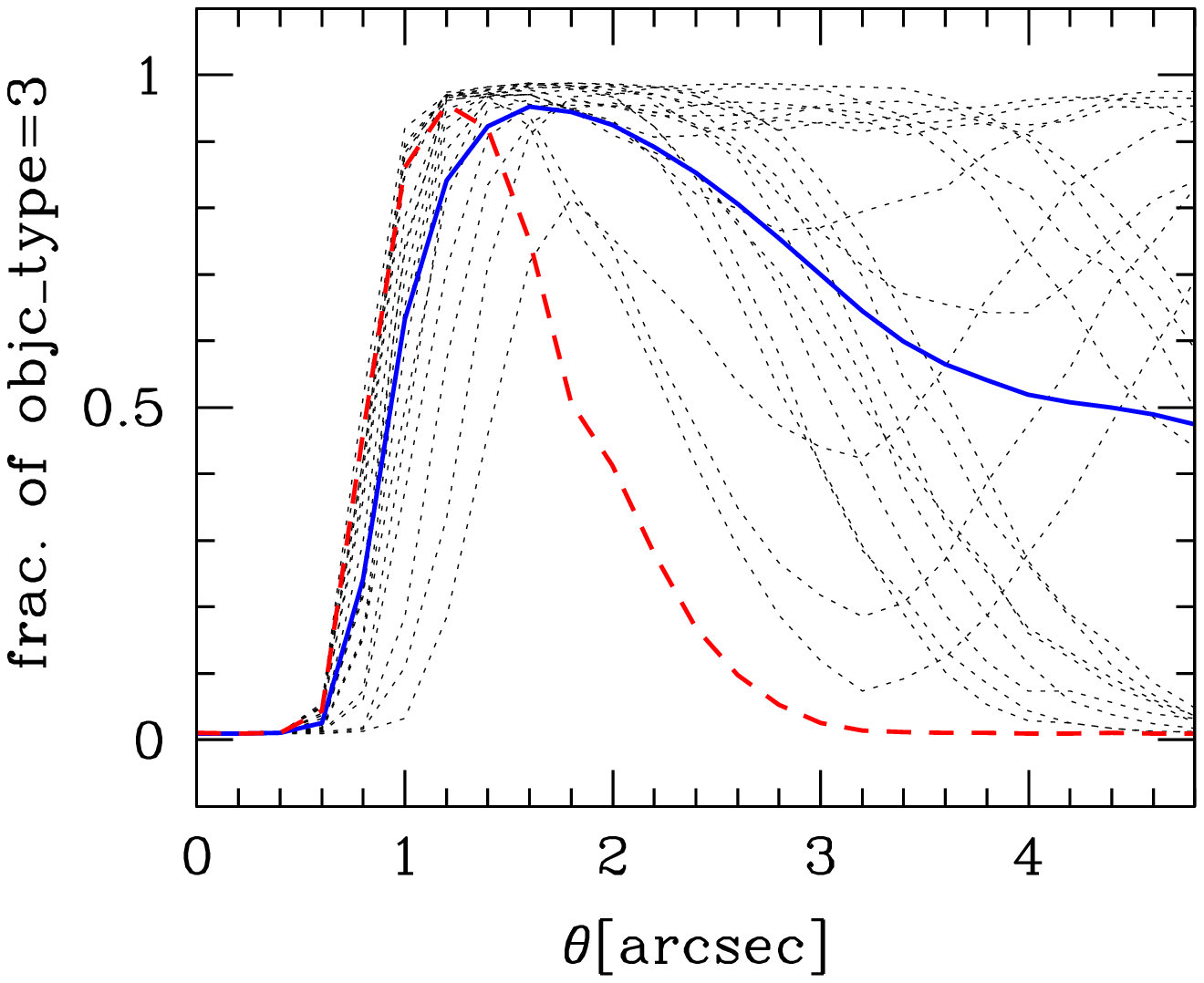}
\caption{The fraction of quadruple lenses that are classified as
  extended objects ({\tt objc\_type} $=3$). The dotted line shows the
  fraction of each quadruple lens image configuration and flux ratios
  taken from real observational data (see Table \ref{table:quad}),
  while the thick solid line indicates the fraction averaged over the
  20 dotted lines. The fraction for double lenses, averaged over the
  flux ratios in the range $10^{-0.5}<f_i<1$, is also shown by a
  dashed line for reference. 
\label{fig:quad_obj}}
\end{figure}

Figure \ref{fig:cont_obj} shows the fraction of double lenses that are
classified as extended objects in the $\theta$-$f_i$ plane. As
expected, the fraction is a strong function of the image separation:
The lenses with $\theta\sim 1\farcs2$ are most likely judged as
extended objects. The fraction depends on the flux ratio; lenses with
larger $f_i$ are more likely to be classified as extended objects. In
Figure \ref{fig:quad_obj} we plot the same fraction for quadruple
lenses. Although the system-by-system difference is large,  at
$\theta\gtrsim 1\farcs5$ quadruple lenses are more likely to be
classified as extended objects on average than double lenses, which is
clearly because the existence of close fold/cusp image pairs.

\end{appendix}


\end{document}